\documentstyle[preprint,aps,psfig]{revtex}
\begin{document}
\draft

\title{Excitonic Strings in one dimensional organic compounds}

\author{St\'{e}phane Pleutin}
\address{Max-Planck-Institut f\"ur Physik Komplexer Systeme, N\"othnitzer
Stra\ss e 38, D-01187 Dresden}

\date{\today}
\maketitle 

\begin{abstract}
Important questions concern the existence of excitonic strings in
organic compounds and their signatures in the photophysics of these
systems. A model in terms of Hard Core Bosons is proposed to
study this problem in one dimension. Mainly the cases with
two and three particles are studied for finite and infinite lattices,
where analytical results are accessible. It is shown that if bi-excitonic
states exist, three-excitonic and
even, {\sl n}-excitonic strings, at least in a certain range of
parameters, will exist. Moreover, the behaviour of the transitions
from one exciton to the biexciton is fully
clarified. The results are in agreement with exact finite cluster
diagonalizations of several model Hamiltonians.
\end{abstract}

\vskip2pc
\narrowtext
\section{introduction}
Excitonic molecules - or biexcitons - are known and well studied since
several decades in conventional semiconductors. Their formations due to
the coupling of a pair of exciton and their photophysics properties
have been well established both theoretically and experimentally
\cite{sc}. The outstanding facts are related to the giant oscillator
strength associated with the one exciton-biexciton transition; we may cite,
for instance, the giant two photon absorption, shown theoretically
by Hanamura \cite{hanamura} and observed first on CuCl by Gale and
Mysyrowicz \cite{gale}. There, five or six orders of magnitude more than for
typical two-photon interband transitions are observed.

In organic compounds, the interest for excitonic bound states - or
excitonic complexes or last, to use a terminology coming from the
Bethe Ansatz technique, excitonic strings - is rather new since the
main studies started about five or six years ago. They concern three
types of compounds: some organic Charge Transfer solids (CT), {\sl J}
or {\sl H}-aggregates ({\sl J-H}A) and Conjugated Polymers (CP).

The organic CT solids considered here, consist of planar aromatic Donor
(D) and Acceptor (A) molecules alternately arranged along a
one-dimensional stack. A large distance between molecules of the same
stack (more than 3 $\AA$) gives a rather small hopping integral
compared to the characteristic Coulomb repulsion. Moreover, a very
large distance between neighbouring chains ensures the quasi-one
dimensionality of these systems. A prototype is given by the anthracene-PMDA
(pyromellitic acid dianhydride); for this compound, by comparison
between theoretical calculations on small clusters and experimental results of
differential transmission spectroscopies, the existence of biexcitons
- and even triexcitons - has been demonstrated recently \cite{string}. This
result stays, for the moment, the only clear evidence for excitonic strings in
organic compounds.

{\sl J} and {\sl H}-aggregates are also, in many cases, one dimensional
stacks of organic molecules. But, on the contrary to the previous class
of compounds, they have no ionic character and the attractive
exciton-exciton interaction is smaller. They are typical organic
solids in the sense that their low-lying excitations can be understood
in terms of Frenkel excitons. These compounds are claimed to
be the ideal ones to get biexcitons but, without convincing
demonstration for the moment - possibly be due to a too small binding energy \cite{frenkel,jh}.

Conjugated polymers are very complex compounds. As far as we are
concerned with low-lying excitations, we may consider them, at first
approximation, as an one dimensional interacting electron gas, which
strongly interacts with the lattice too \cite{revue}. Moreover, for a more
realistic description, in many cases, three dimensional effects and
inter- and intra-chain disorder should be considered. Because of these
difficulties, despite a large amount of work, even the nature of the
primary excitations remains subject of intense controversies
\cite{excitations}. Anyway, some Photoinduced Absorption (PA)
experiments \cite{pa,klimov} and Two Photon Absorption (TPA) experiments
\cite{tpa,abe} show some features which are possibly due to biexcitons. 

Only a few theoretical investigations of excitonic strings in organic
compounds have been done
\cite{string,frenkel,jh,abe,mazumdar,vektaris,spano1,spano2,yu,gallagher}. Most of them are based on 
exact diagonalizations of small clusters using different kinds of model
Hamiltonians depending on the system under consideration \cite{string,frenkel,jh,mazumdar}. From these results and based on physical
reasoning, it has been argued that, when biexcitonic states exist, the
intensity of the optical one exciton-biexciton transition may decrease
when the binding energy of the biexciton increases; moreover, this
intensity is shown to be less important than the one of the one
exciton-two excitons (two free excitons) transitions and, even, is
supposed to be independent of the system size for sufficiently large
binding energies \cite{frenkel}. Therefore, if the tendencies observed
on finite clusters remain the same for the infinite system, on the
contrary to the situation observed in some inorganic semiconductors \cite{hanamura,gale}, it may be
difficult to create biexcitons by TPA. A more appropriated experimental
technique to observe excitonic strings would then be the Differential Transmission
Spectroscopy used by the way in \cite{jh}. Throughout this manuscript, we often
refer to these numerical works for comparisons, especially concerning the
intensity of the transition from one exciton to biexciton states.

In this paper, we propose a model for studying
excitonic complexes in organic compounds. Based on physical arguments, we
think, this model is relevant for the three classes of compounds
enumerated above. For organic crystals, as the $J$ and $H$ aggregates,
our model is equivalent, in the low energy sector, to an effective {\sl XXZ}
spin one-half Heisenberg model already
used to study this problem \cite{frenkel,vektaris,spano1,spano2}. For the CT solids and CP,
some extensions are necessary because the proper excitons are not
of Frenkel type. They
are realized by introducing effective Hard Core Boson particles with
a finite extension on a lattice, which mimics the exciton
extension. Next, we propose an effective Hamiltonian in terms of these
particles which captures the essential physical ingredients to describe
excitonic strings in organic compounds. This model is considerably
easier to study than the traditionally employed models and therefore some
analytical investigations are possible. In this paper, we study
the proposed model for two and three particles in a finite and infinite
lattice. Our main interests concern the behaviours of the binding
energies of the excitonic strings with the parameters of the model and
of the one exciton-biexciton transition oscillator strengths; are they
giant as in inorganic semiconductors or, on the contrary, small as suggested by
exact diagonalizations on small clusters \cite{frenkel,mazumdar}? This
question is crucial in order to clarify the role of hypothetical
biexcitons - or even more extended excitonic strings - in the
photophysics of organic compounds. 

The paper is organized as follow. In section II, we introduce our Hard Core Boson
Hamiltonian. In section III, IV and V, we study this Hamiltonian for one, two
and three particles respectively.

\section{model}

First, for each class of systems mentioned in this study, we emphasize
the main characteristics which are important for our purpose.

{\sl J} and {\sl H} aggregates are typical Organic Crystals. The
ground state of these systems is the tensorial product of the
molecular ground states. The low-lying excitations are
Frenkel-type excitons where the hole and the electron of a monoexcitation are located on the same
molecular site \cite{frenkel,vektaris,spano1,spano2,davydov}. We can draw a picture of these
excitations in the following way: if we
represent symbolically the ground state of such a
quasi-one dimensional system as $...MMMMMM...$, where $M$ stands for a Molecule in its ground state,
the Frenkel excitons are the Bloch states of local configurations such
as $...MMM^{\pm}MMM...$ where the superscripts $-$ and $+$ are for the
electron and the hole, respectively; $M^{\pm}$ is then a monoexcited
molecule.

Organic CT solids are stacks of Donor (D) and Acceptor (A) molecules. We
already mentioned that they are, with a good approximation, quasi-one
dimensional systems with a nearest-neighbour hopping term
small compared to the Coulombic term. The strong coupling regime is then a
good starting point to consider such systems. Moreover, if one consider
only a short range Coulombic term, their
ground state would be a Charge Density Wave depicted
roughly by the unique configuration $...DADADA...$ where D molecules
get two electrons and A molecules none. The low-lying excitations
would then be the excitons where an electron is transfered from a D to
an A
molecule; they would be Bloch states of excited configurations such as
$...DAD^{+}A^{-}DA...$ (more details
could be found in ref \cite{string}). We get a so-called Charge Transfer Exciton
(CTE). For a more realistic Coulomb potential, the picture proposed
above will remain reasonable, at first approximation.

Conjugated Polymers are quite different from the other
systems just considered since, obviously, they are not molecular
crystals. On the contrary, they are usually
described as an one dimensional electron gas governed by the
Pariser-Parr-Pople (PPP) Hamiltonian which includes both long range Coulomb
interaction and, semi-classically, electron-phonon interaction
\cite{revue}. Numerical studies of this Hamiltonian give, for a proper
choice of parameters, excitonic states of small radius, typically a few
monomeric units \cite{abe,yu}. On the other hand, the ground state may be
analyzed in terms of basis sets completely localized on the
monomers. Then, the ground state of finite
clusters is shown to contain mainly intermonomer charge fluctuations of
very short range - especially the nearest-neighbours ones
\cite{chandross}. According to these specificities, an
effective model has been proposed very recently \cite{pleutin1,pleutin2}. Starting
from the PPP Hamiltonian, this model reaches a kind of
molecular-crystal description for
the low-lying excitations in conjugated polymers. In this new approach the
vacuum (or ground state) is still not the simple molecular-type of
ground state, proper to the Simpson related models \cite{simpson,rice}, but a
matrix-product like state composed by local configurations extended at
most, over two monomers. Excitons are then composed mainly by local
configurations where either the electron and the hole are localized on
the same monomer or distant by only one monomer; the weight of the
configurations where the electron and the hole are more separated decreases exponentially in the
wave functions. Therefore, the radius of the exciton, $r$, defined as an average distance between the hole and the electron is  rather small. If we assume
$r=1$, the following
pictorial representation can be done: if we
represent the ground state as $...MMMMMM...$, where
$M$ is a Monomer, the excitons will be the Bloch
states of such configurations
$(...MMM^{+}M^{-}MM...)\pm(...MMM^{-}M^{+}MM...)$. Because of
the electron-hole symmetry, the charge transfer of the electron on the
right side and on the left side must be combined either symmetrically or
antisymmetrically. In contrast to the two previous cases, excitons
have here, a structure. 

In conclusion, in these three cases, excitons are Bloch
states of very localized excitations. The vacuum and the nature of these local
excitations vary, of course, following the system under consideration:
crystal molecular ground state and Frenkel
excitons for $J$ and $H$ aggregates, CDW and CTE for CT solids and
finally, matrix-product state and some kind of CTE but
including the two possible arrangements of the charges, for
CP. However, all these very localized excitations can be thought in
terms of effective Hard Core Bosons (HCB) extended over a few bonds -
with an extension
of the order of the exciton's radius. We adopt such a view in
this work, and describe the dynamics of
these excitations within the same model which we
introduce now.

In this study, we consider only one dimensional lattices with N
sites, a site being a molecular-site in the case of molecular
crystals\cite{string} and a monomer-site in the case of CP
\cite{pleutin1,pleutin2}. As we already mentioned, this is a good
approximation for a large class of compounds studied here. However, the generalization to higher dimensions should be
possible and the subject of further studies.

Excitons are extended over $r$ sites, the radius of the excitons, which varies from zero to
a few units in the cases considered here. In this study, to mimic the excitons, we
introduce HCB particles extended over one bond ($r=1$). This 'minimal'
extension is adopted more by a pictorial preoccupation but, also, in
accordance with the
work developed in ref \cite{pleutin1,pleutin2}. Indeed, at the thermodynamic limit where we will
derive all the analytical formulae, this extension plays obviously no
role, since it remains small compared to the system size. Moreover,
for finite size cluster calculations, the qualitative behaviours shown
in this paper could also be obtained by choosing $r=0$, $r=2$ or even
other 'small' $r$.  The restriction to $r=1$ does not
affect the conclusions but, if necessary, more (or less) extended
particles could also be considered without any difficulties. Working
with extended particles, instead of particles localized on a site, does
not imply technical complications.

\begin{figure}
\centerline{\psfig{figure=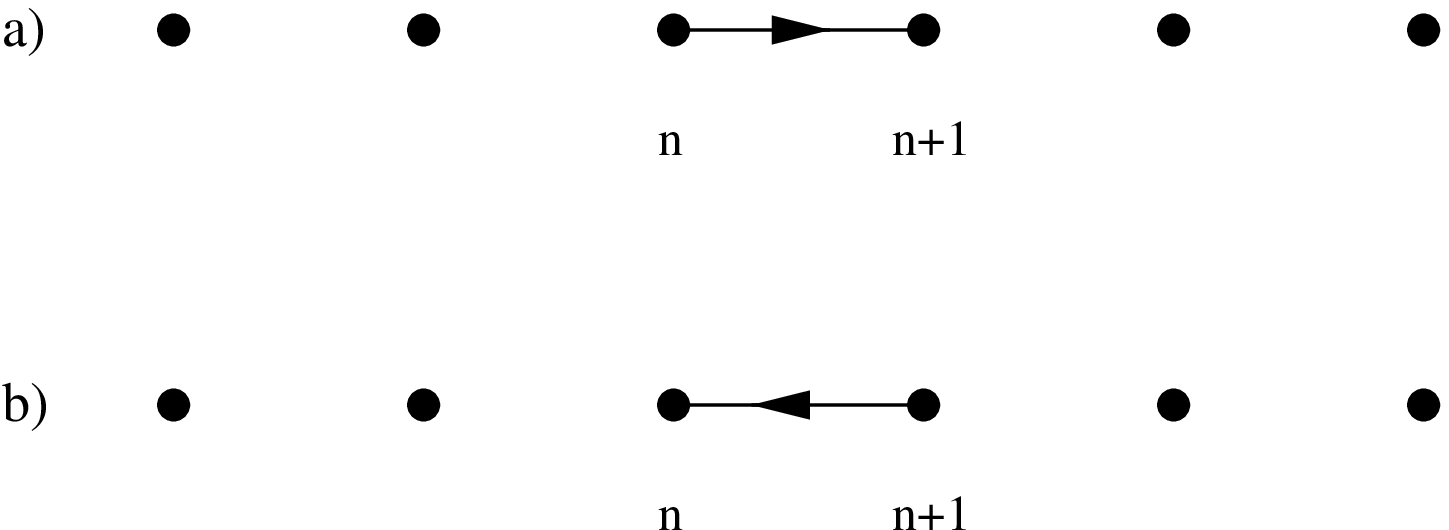,width=8cm}}
\caption{Pictures for a) Right-bosons and b) Left-bosons. The arrows
show the position of the electron, the hole being on the other side}
\label{RLbosons}
\end{figure}

The Coulomb interaction
between excitons must be included in order to describe excitonic strings . For that purpose, the relative
position between the electron and the hole, the two elementary
constituents of an exciton, becomes important. We
introduce then two species of HCB extended over one
bond:
\begin{itemize}
\item The so-called Right-bosons ($R$) which mimic nearest
neighbour electron-hole pairs with the electron on the right (see figure \ref{RLbosons}.a).
\item The so-called Left-bosons ($L$) which also mimic
nearest neighbour electron-hole pairs but with the electron on the
left (see figure \ref{RLbosons}.b).
\end{itemize}

The $R$ and $L$ particles obey the following Hamiltonian

\begin{equation}
\label{hamiltonian}
\begin{array}{c}
H=\omega_{0}\sum_{n}(R_{n}^{\dagger}R_{n}^{\quad}+L_{n}^{\dagger}L_{n}^{\quad})-J
\sum_{n}(R_{n}^{\dagger}R_{n+1}^{\quad}+h.c.)-J\sum_{n}(L_{n}^{\dagger}L_{n+1}^{\quad}+h.c.)\\-
\alpha
\sum_{n}(R_{n}^{\dagger}L_{n}^{\quad}+h.c.)-V\sum_{n}(\hat{N}^{R}_{n}-\hat{N}^{L}_{n})(\hat{N}^{R}_{n+2}-\hat{N}^{L}_{n+2})
\end{array}
\end{equation}$R_{n}^{(\dagger)}$ and $L_{n}^{(\dagger)}$ are the
destruction (creation) operators of the $R$ and $L$ bosons at site
$n$, $\hat{N}^{R(L)}_{n}$, the operators number of HCB $R$ ($L$) at
site $n$. $\omega_{0}$ is the excitonic energy ($\omega_{0}>0$). The $J$
term is the hopping term of the HCB ($J>0$); it is directly related to
the exciton band width given by $4J$. $\alpha$ is an effective term
for virtual interactions which couple locally $R$ and $L$ bosons
($\alpha>0$). The $R$ and $L$ particles are effective representations
of more complex particles, invoking a more sophisticated level of
description; these "high level" particles possess a structure which
can be very complicated as it is the case for most of the CP. The interaction
$\alpha$ reflects the existence of virtual processes exchanging the
relative position of the electron and the hole of an exciton; this can
be due, for instance, to kinetic terms. Indeed, by a second order process
involving a n.n. hopping term, it is readily
possible to exchange the relative position of the two building particles of an
exciton. The last term of (\ref{hamiltonian}), $V$ ($V>0$), is the
interaction between excitons restricted to nearest-neighbours. We have
to distinguish between two situations: for two excitons of the same
species, $RR$ or $LL$, the nearest charges are of opposite sign so
that the resulting interaction must be attractive; the situation is
reversed for two excitons of opposite species, $RL$ or $LR$, where the
resulting interaction is repulsive.

For structureless excitons, Frenkel or CT-excitons, the definition of
two species of degenerate HCB makes no sense. To study
strings of such excitons with the general Hamiltonian
(\ref{hamiltonian}), it is sufficient to: (i) identify the $R$ and $L$
particles and (ii) to take the limit $\alpha=0$. To be more precise,
$R$ and $L$ particles may exist in the case of CT solids, but with a
large energy difference which results, in our model, in a very small
$\alpha$; therefore, it is a good approximation to simply neglect this
term in this case.

In the following, we will study the Hamiltonian (\ref{hamiltonian})
for one, two and three particles, which are the cases of experimental interest.

\section{One exciton states}
Excitons appear as Bloch states
of symmetric or antisymmetric combinations of $R$ and $L$ bosons,
corresponding to two different symmetry classes

\begin{equation}
\label{1exciton}
\mid
\pm,k>=\frac{1}{\sqrt{2(N-1)}}\sum_{n}e^{ikn}(R_{n}^{\dagger}\pm
L_{n}^{\dagger})\mid0>
\end{equation}where $\mid0>$ is the state without any boson. The
separation in energy between this two states is given by $2\alpha$,
the $(+)$ state being lower.

In the case of {\sl J} ({\sl H}) aggregates and CT solids, we identify
$R$ and $L$ bosons and get only one excitonic state
\begin{equation}
\mid -,k>=\mid0>\quad,\quad \mid +,k>=\mid k>=\frac{1}{\sqrt{N-1}}\sum_{n}e^{ikn}R_{n}^{\dagger}\mid0>
\end{equation}

In all cases, exciton states result in a very intense peak in the linear
absorption spectrum for $k=0$, while the electron-hole continuum is almost not
visible. This very unusual feature, if we think about
conventional semiconductors, is an important characteristic of these
one dimensional compounds \cite{pleutin1,pleutin2,sylvie}. This is
easily understandable since the radius of the exciton is very small
and the ground state contains mainly short range charge
fluctuations \cite{pleutin1,pleutin2}.

  Now, we write down the elementary transition
moments due to a localized HCB, $\vec{m}_{cp}$, in the case of CP
and $\vec{m}_{oc}$, in the case of {\sl J} ({\sl H}) aggregates and CT
solids ($oc$ is for Organic Crystal);
these quantities will serve in the
following to express the transition moments between one exciton and
two exciton states

\begin{equation}
\label{moment}
\begin{array}{l}
\vec{m}_{cp}=\frac{1}{\sqrt{2}}<0\mid
e\vec{r}(R_{n}^{\dagger}+L_{n}^{\dagger})\mid0>\\
\vec{m}_{oc}=<0\mid e\vec{r}R_{n}^{\dagger}\mid0>
\end{array}
\end{equation}where $e$ is the charge of the electron and $\vec{r}$ the
position operator.

The intensity of the ground state-one exciton
transition at $k=0$ is proportional to $N \mid \vec{m}_{oc} \mid^{2}$ or $N
\mid \vec{m}_{cp} \mid^{2}$, depending on the system under
consideration \cite{pleutin2}. In the case of CP, only the excitonic
state $\mid +,0>$ is observable in linear absorption. However, the
state $\mid -,0>$ could become important, as it is the case in
electroabsorption experiments for  instance.

\section{Two exciton states}
The wave functions with two excitons are written as
\begin{equation}
\mid
\Psi_{2}>=\sum_{n_{1}<n_{2}-1}\Psi_{\sigma_{1}\sigma_{2}}(n_{1},n_{2})\mid
n_{1},n_{2}>
\end{equation}where $\sigma_{1}$ and $\sigma_{2}$ are $R$ or $L$
bosons, $\mid n_{1},n_{2}>$ the ket relative to the situation where
the particle $\sigma_{1}$ is in $n_{1}$ and $\sigma_{2}$ in
$n_{2}$ with $n_{1}<n_{2}-1$,
$\Psi_{\sigma_{1}\sigma_{2}}(n_{1},n_{2})$ is the corresponding
amplitude.

As usual, we treat separately the movement of the center of mass
and of the relative position of the two particles. Moreover, we have to
specify the nature, $R$ or $L$, of the particles located on
$n_{1}$ and $n_{2}$, it is then convenient to use matrix
formalism. We write
\begin{equation}
\label{wf}
\mid \Psi_{2}>=\sum_{n>1,n_{1}}e^{iQ(n_{1}+n_{2})}[\varphi]_{n}\mid n_{1},n_{2}>
\end{equation}with $n=n_{2}-n_{1}$ ($n>1$) and

\begin{equation}
[\varphi]_{n}=\left(
\begin{array}{l}\varphi_{rr}(n)\\ \varphi_{rl}(n)\\ \varphi_{ll}(n)\\ \varphi_{lr}(n)
\end{array}\right)
\end{equation}where the small left (right) indices refer to the nature
of the left (right) particle, $R$ or $L$ bosons.

We apply Periodic Boundary Conditions (PBC) to the wave functions which give
the following constraints

\begin{equation}
\left\{ \begin{array}{l} Q=\frac{j2\pi}{N} \quad,\quad j=1,...,N \quad
(j \, \mbox{an integer})\\ 
\left[\varphi\right]_{-n+N}=[\varphi]_{n} \quad \mbox{where} \quad n=n_{2}-n_{1}
\end{array}\right.
\end{equation}

Moreover, we work with Hard Core Bosons extended over one link so
that we must preserve
\begin{equation}
[\varphi]_{0}=[\varphi]_{1}=[0]
\end{equation}where $[0]$ is the zeroth four component vector.

With the wave function (\ref{wf}), the eigenvalue equation is written
\begin{equation}
\label{vp}
E[\varphi]_{n}=-2J\cos Q[I]\{[\varphi]_{n-1}+[\varphi]_{n+1}\}-\alpha[\Sigma][\varphi]_{n}-\delta_{n,2}V[\Lambda][\varphi]_{n}
\end{equation}where $[I]$ is the 4 by 4 identity matrix, and
\begin{equation}
[\Sigma]=\left( \begin{array}{c}
 0 \quad 1 \quad 0 \quad 1 \\ 1 \quad 0 \quad 1 \quad 0
\\ 0 \quad 1 \quad 0 \quad 1 \\ 1 \quad 0 \quad 1 \quad 0
\end{array}\right) \quad , \quad 
[\Lambda]=\left(
\begin{array}{c}
1 \quad \quad \quad \quad \quad \quad \\
\quad \quad -1 \quad \quad \quad \quad\\ \quad \quad \quad \quad 1
\quad \quad \\ \quad \quad \quad \quad \quad \quad -1
\end{array}\right)
\end{equation}

With the equation(\ref{vp}), we have reached an impurity like problem
where free particles, belonging to four different continuum, interact
with impurities pinned at $n=2$. First, by applying a local unitary
transformation, we solve (\ref{vp}) away from the impurities.
\begin{equation}
\label{unitary}
\begin{array}{c}
[\bar{\varphi}]_{n}=[U][\varphi]_{n}\\ \quad \\
\left(\begin{array}{c}\varphi_{0}(n)\\ \varphi_{\bar{0}}(n)\\
\varphi_{+}(n)\\ \varphi_{-}(n)\end{array}\right) = 
\left(\begin{array}{crlrlrl} 
\frac{1}{\sqrt{2}} &    & 0                 & -  & \frac{1}{\sqrt{2}} &
& 0 \\ 
0                  &    & \frac{1}{\sqrt{2}}&   & 0                  &
-  & \frac{1}{\sqrt{2}} \\ 
\frac{1}{2}        &    & \frac{1}{2}       &   & \frac{1}{2}        &
& \frac{1}{2} \\ 
\frac{1}{2}        & -  &  \frac{1}{2}      &   & \frac{1}{2}        &
- & \frac{1}{2} 
\end{array}\right)

\left(\begin{array}{c}\varphi_{rr}(n)\\ \varphi_{rl}(n)\\ \varphi_{ll}(n)\\ \varphi_{lr}(n)
\end{array}\right)
\end{array}
\end{equation}

We can now rewrite equation (\ref{vp}) in terms of
$[\bar{\varphi}]_{n}$; we find
\begin{equation}
\label{vp2}
E[\bar{\varphi}]_{n}=-2J\cos Q[I]\{[\bar{\varphi}]_{n-1}+[\bar{\varphi}]_{n+1}\}-2\alpha[\sigma_{2}][\bar{\varphi}]_{n}-\delta_{n,2}V\{[\sigma_{1}][\bar{\varphi}]_{n}+[\sigma_{3}][\bar{\varphi}]_{n}\}
\end{equation}where
\begin{equation}
[\sigma_{1}]=\left(\begin{array}{c}1 \quad \quad \quad \quad \quad
\quad \\ \quad \quad -1 \quad \quad \quad \quad \\ \quad \quad \quad
\quad  0 \quad \quad \\ \quad \quad \quad \quad \quad \quad
0\end{array}\right), \quad [\sigma_{2}]=\left(\begin{array}{c} 0
\quad \quad \quad \quad \quad \quad \\ \quad \quad 0 \quad \quad \quad
\quad \\ \quad \quad \quad \quad 1 \quad \quad \\ \quad \quad  \quad
\quad \quad \quad -1\end{array}\right), \quad
[\sigma_{3}]=\left(\begin{array}{c}0 \quad \quad \quad \quad \quad
\quad \\ \quad \quad 0 \quad \quad \quad \quad \\ \quad \quad  \quad
\quad 0 \quad 1 \\ \quad \quad \quad \quad 1 \quad 0\end{array}\right)
\end{equation}

Without the impurity like states at $n=2$, there is no mixing between
$\varphi_{0}$, $\varphi_{\bar{0}}$, $\varphi_{+}$ and
$\varphi_{-}$ components; with the scattering of particles at $n=2$,
only the
$\varphi_{+}$ and $\varphi_{-}$ components are mixed.

From equations (\ref{moment}) and (\ref{unitary}), it
is clear that only the $\varphi_{+}(n)$ and the $\varphi_{-}(n)$
components of $[\bar{\varphi}]_{n}$ are
relevant to study the optical responses of  two exciton states. Since we are mainly
interested in spectroscopic properties of biexcitons, we consider only these two components in
the following. The $\varphi_{0}(n)$ and $\varphi_{\bar{0}}(n)$
components have no physical meaning for structureless excitons, but
could be important for CP; their characteristics could be calculated,
following exactly the same way as the one presented here.

We write, following H. Bethe \cite{bethe},
\begin{equation}
\label{fonctions}
\varphi_{+}(n)=A_{+}e^{\mu n}+B_{+}e^{-\mu n}\quad \mbox{and} \quad \varphi_{-}(n)=A_{-}e^{\eta n}+B_{-}e^{-\eta n}
\end{equation}which give the following expressions for the energy where
$J(Q)=2J\cos Q$ and where we choose $2\omega_{0}$ as reference
\begin{equation}
\label{energie}
E=-2\alpha-J(Q)(e^{\mu}+e^{-\mu})\quad \mbox{and} \quad
E=2\alpha-J(Q)(e^{\eta}+e^{-\eta})
\end{equation}

$A_{+}$, $A_{-}$, $B_{+}$ and $B_{-}$, $\mu$ and $\eta$ are the
constants to be determined. Note that, if in (\ref{fonctions}) $\mu$
and $\eta$ are purely imaginary numbers,
we get free excitons. On the contrary, if they are purely real numbers,
we get bound-states, the biexcitons.

The ansatz (\ref{fonctions}) gives obviously the exact solution of (\ref{vp2})
without impurities ($V=0$); with impurities, (\ref{fonctions}) must fulfill the
following constraints in order to remain solution of (\ref{vp2}) \cite{bethe}
\begin{equation}
\label{impurities}
\begin{array}{c}J(Q)\varphi_{+}(1)-V\varphi_{-}(2)=0\\
J(Q)\varphi_{-}(1)-V\varphi_{+}(2)=0
\end{array}
\end{equation}

Moreover, the PBC imply
\begin{equation}
\label{PBC}
A_{+}=e^{-\mu N}B_{+}\quad and \quad A_{-}=e^{-\eta N}B_{-}
\end{equation}

In the following, we give the solutions of the above mentioned problem for
different choices of parameters.
 
\subsection{$\alpha=0$, $V=0$}
In this trivial case, the two excitons are free; we get two degenerated continuum for
$\varphi_{+}(n)$ and $\varphi_{-}(n)$ within the energy range
$[-2J(Q),2J(Q)]$. It is easy to show that a very intense linear
absorption takes place for $Q=0$ and $\mu=\eta=0$ only; then the transition moment of the
one exciton - two free exciton transition at the edge of the continuum behaves as

\begin{equation}
\label{momentfree}
\mid \vec{M}_{1\ exciton-2\ free\ excitons} \mid^{2}\sim N\mid
\vec{m}_{cp/oc} \mid^{2}
\end{equation}plus corrections in $\frac{1}{N}$. For $\alpha \ne 0$
and/or $V \ne0$, this strong absorption remains but slightly shifted
toward the high energies.

This transition is similar to the ground state - one exciton
transition; indeed, the excitation energies and the behaviour of the
transition moments are the same in these two cases. This
characteristic was already used in \cite{string,frenkel} to explain the lack
of bleaching signal in PA experiments for Charge Transfer and Frenkel exciton systems.

\subsection{$\alpha=0$, $V\neq0$}
This case concerns CT solids and {\sl J} or {\sl H} aggregates for
instance. Without the $\alpha$ term, our model is equivalent to a
standard interacting Frenkel exciton Hamiltonian already studied
numerically \cite{frenkel} and analytically
\cite{vektaris,spano1,spano2}. Our results, obtained with a different
technique, are in accordance with the conclusions reach by these
various studies.

Since $V\neq0$, bound states - or biexcitons - may exist; in
this subsection we consider the biexcitonic states only and then
$\mu$ and $\eta$ of equations (\ref{fonctions}), as real quantities.

When $\alpha=0$, equations (\ref{energie}) imply that
$\mu=\eta$. From equations (\ref{impurities}) together with the
PBC (\ref{PBC}), we get at the thermodynamic limit, the following set
of equations
\begin{equation}
\label{set}
\begin{array}{l}
B_{-}=\lambda B_{+}e^{\mu}\\
\lambda^{2}-e^{-2\mu}=0
\end{array}
\end{equation}where $\lambda=\frac{J(Q)}{V}$.

Additionally to the continuum, if $\lambda<1$, i.e $V>J(Q)=2J\cos Q$, we get with the second
equation of (\ref{set}) and equations (\ref{energie}), two bound states with energies given by the
following simple expressions
\begin{equation}
\label{energiebiexciton}
E=-V-\frac{J^{2}(Q)}{V}\quad \mbox{and} \quad E=+V+\frac{J^{2}(Q)}{V}
\end{equation}
By using a two-particle Green function approach \cite{vektaris}, these
equations have already been obtained for $Q=0$ in \cite{spano1} and
for every $Q$ in \cite{vektaris,spano2} together with the biexciton
condition $V>J(Q)$. 

The first equation gives a bound state below the continuum of two free
exciton states, this is the biexciton of interest for us; the second equation gives another
bound state on top of the continuum. In the following, we consider
only the first state and calculate its wave function and
the transition moment associated with the one exciton-biexciton
transition. 

First, we write $A_{\pm}=e^{-\delta/2}$ and
$B_{\pm}=e^{\delta/2}$. With the PBC (\ref{PBC}) and equations (\ref{set}), we get the
expression for $\delta$
\begin{equation}
-\delta=N \ln \lambda
\end{equation}Then, with equations (\ref{fonctions}) and (\ref{set}), we get
\begin{equation}
\varphi_{\pm}={\cal N} \cosh(-\ln{\lambda}[n-\frac{N}{2}])
\end{equation}valid in the interval $1<n<N-1$. ${\cal N}$ is the
normalization constant given by
\begin{equation}
{\cal N}=\left[\frac{1}{2}(N-3)+\frac{1}{2(1-\lambda^{2})}\left(\frac{1}{\lambda^{N-4}}-\lambda^{N-2}\right)\right]^{-\frac{1}{2}}
\end{equation}

The first equation of (\ref{set}) shows that the wave function of the
lowest bound state is a symmetric mixture of the $\varphi_{+}$ and
$\varphi_{-}$ components resulting in configurations with two like
particles $RR$ or $LL$. On the contrary, the highest bound state
contains only configurations with two unlike particles $RL$ or
$LR$. We recall that for this case $R$ and $L$ particles are identified
so that the distinction between the case with $RR$ or $LL$ bosons and
$RL$ or $LR$ bosons is then a way to treat the case
of attractive and repulsive interaction between bosons at the same time. In case of
attractive interaction (two bosons of same
species), we get  a bound state below the two exciton continuum; the
situation is reversed in the case of repulsive interaction (two bosons
of opposite species), where a bound state appears above the two
exciton continuum \cite{frenkel}.

For large systems, the transition moment of the one exciton-biexciton
transition is non-zero only for $Q=0$ and $k=0$ ($Q$, the center of
mass of the biexciton, $k$, the exciton momentum); its expression is then given
with the help of the elementary transition moment (\ref{moment}) by (more
precisely, the square of the transition moment is shown here)
\begin{equation}
\label{momentN}
\mid \vec{M_{oc}}\mid^{2}=2\mid \vec{m}_{oc}\sum_{n}\varphi_{+}(n)\mid^{2}=2\mid
\vec{m}_{oc}\mid^{2}{\cal
N}^{2}\left(\frac{1}{1-\lambda}\left(\frac{1}{\lambda^{N/2-2}}-\lambda^{N/2-1} \right) 
\right)^{2}
\end{equation}Going to the thermodynamic limit, we found the simple result
\begin{equation}
\label{momentthermo}
\mid \vec{M_{oc}}\mid^{2}= 4\mid\vec{m_{oc}}\mid^{2}\frac{1+\lambda}{1-\lambda}
\end{equation}which is exactly the result found in \cite{spano1,spano2} by
using a two particle Green function approach.

We recall that the bound states (biexcitons) exist only if $0<\lambda<1$
and that the biexciton binding energy increases when $\lambda$
approaches $0$. With this formula, the behaviour of the absorption
from the one exciton to biexciton state appears very clearly. First les us
consider its behavior when $\lambda$ approaches $0$ or $1$, at the
thermodynamic limit (equation (\ref{momentthermo})): for $\lambda
\rightarrow 0$, the transition moment reaches saturation to
$4\mid\vec{m_{oc}}\mid^{2}$ which is easily understandable since, in this
case, the two excitons are strongly bounded together in two n.n. sites;
for $\lambda \rightarrow 1$, the transition moment diverges following the asymptotic behaviour in
$4\mid\vec{m_{oc}}\mid^{2}\frac{2}{1-\lambda}$. Next, we can make some
conclusions by studying the equation (\ref{momentN}).
\begin{itemize}
\item At large and fixed $N$, with decreasing $\lambda$ (which means, with
increasing the biexciton binding energy), the intensity of the
transition decreases.
\item At fixed $\lambda$ and sufficiently large $N$, $\vec{M_{oc}}$ is
almost independent of the system size. On the contrary, the intensity of the transition between one exciton
and two free excitons increases linearly with the system size
(cf. eq. \ref{momentfree}). However, as expected, when $\lambda$ approaches $1$,
$\mid \vec{M_{oc}}\mid^{2}$ recovers progressively a linear dependence
in $N$.
\end{itemize}

Both of these behaviours depict inverse mechanisms than the ones
theoretically predicted for the ground state-one exciton transition
\cite{pleutin1,pleutin2}, where the intensity of the excitonic peak
is proportional to the size of the system and increases with the
binding energy.

\begin{figure}
\centerline{\psfig{figure=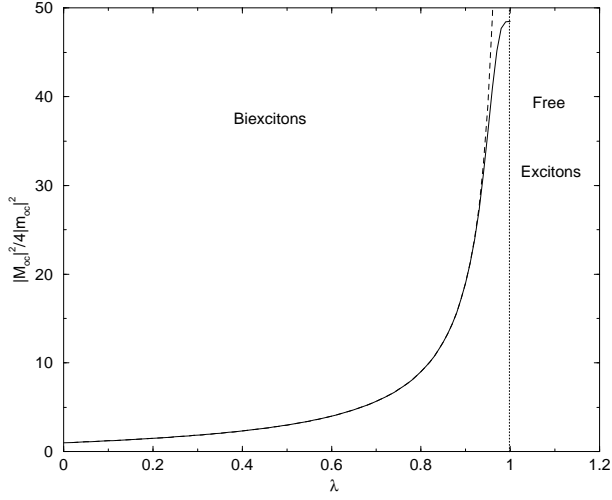,width=8cm,angle=-90}}
\caption{Intensity of the one exciton-biexciton transition for N=100 (full
line) and at the thermodynamic limit (dashed line) as
a function of $\lambda=\frac{2J}{V}$ ($\alpha=0$, $Q=0$)}
\label{intensity}
\end{figure}

These behaviours are illustrated in figure \ref{intensity} where the equation
(\ref{momentN}) is evaluated for
$N=100$. There, we see that the intensity of the one exciton-biexciton
transition decreases dramatically when $\lambda$ decreases (which
means, when the binding energy increases) reaching saturation for small $\lambda$. Then, this intensity is independent of
$N$ which gives in PA and TPA spectroscopy an intensity proportional to
$\sqrt{N}$ instead of $N$ for the transition to the continuum of two
free excitons; this last behaviour can be seen by comparison with equation
(\ref{momentthermo}) evaluated at the large $N$ limit (dashed line). Consequently, at the thermodynamic limit, biexcitons
may be observable by spectroscopy experiments, but with the need of very clean
compounds, otherwise we may expect at first analyse, that the intense
transition to two free excitons would not permit to detect possible biexcitons. Within our model, no two-photon
giant resonance \cite{hanamura} can be
expected for $J$, $H$ aggregates and organic Charge Transfer Solids. The very same conclusions were already pointed out in
ref \cite{frenkel} from numerical calculations and in
\cite{spano1,spano2} from analytical studies.

\subsection{$\alpha\neq0$, $V\neq0$} 

This case concerns CP. As in the previous subsection, we look
for bound states only so that $\mu$ and $\eta$ are again assumed to be
real quantities. The question we want to address more
specifically here is how
the $\alpha$ term will affect the binding energy, the wave function
and the transition moment related to the biexcitons. 

For $\alpha \neq0$, from equations (\ref{impurities}) and (\ref{PBC}),
we get at the thermodynamic limit the following set of equations

\begin{equation}
\label{set2}
\begin{array}{l}
B_{-}=\lambda e^{-\mu}e^{2\eta}B_{+}\\
\lambda^{2}e^{\eta}-e^{-\mu}=0
\end{array}
\end{equation}

From equations (\ref{energie}) and the second equation of
(\ref{set2}), we get
\begin{equation}
\label{expeta}
e^{-\mu}=\frac{\lambda^{2}}{1-\lambda^{2}}\left(\beta+\sqrt{\beta^{2}-(1-\lambda^{2})(1-\frac{1}{\lambda^{2}})}\right)
\end{equation}where $\lambda=\frac{J(Q)}{V}$ and
$\beta=\frac{\alpha}{J(Q)}$. This expression together with equations
(\ref{energie}), gives the energy of the lowest biexciton.

\begin{figure}
\centerline{\psfig{figure=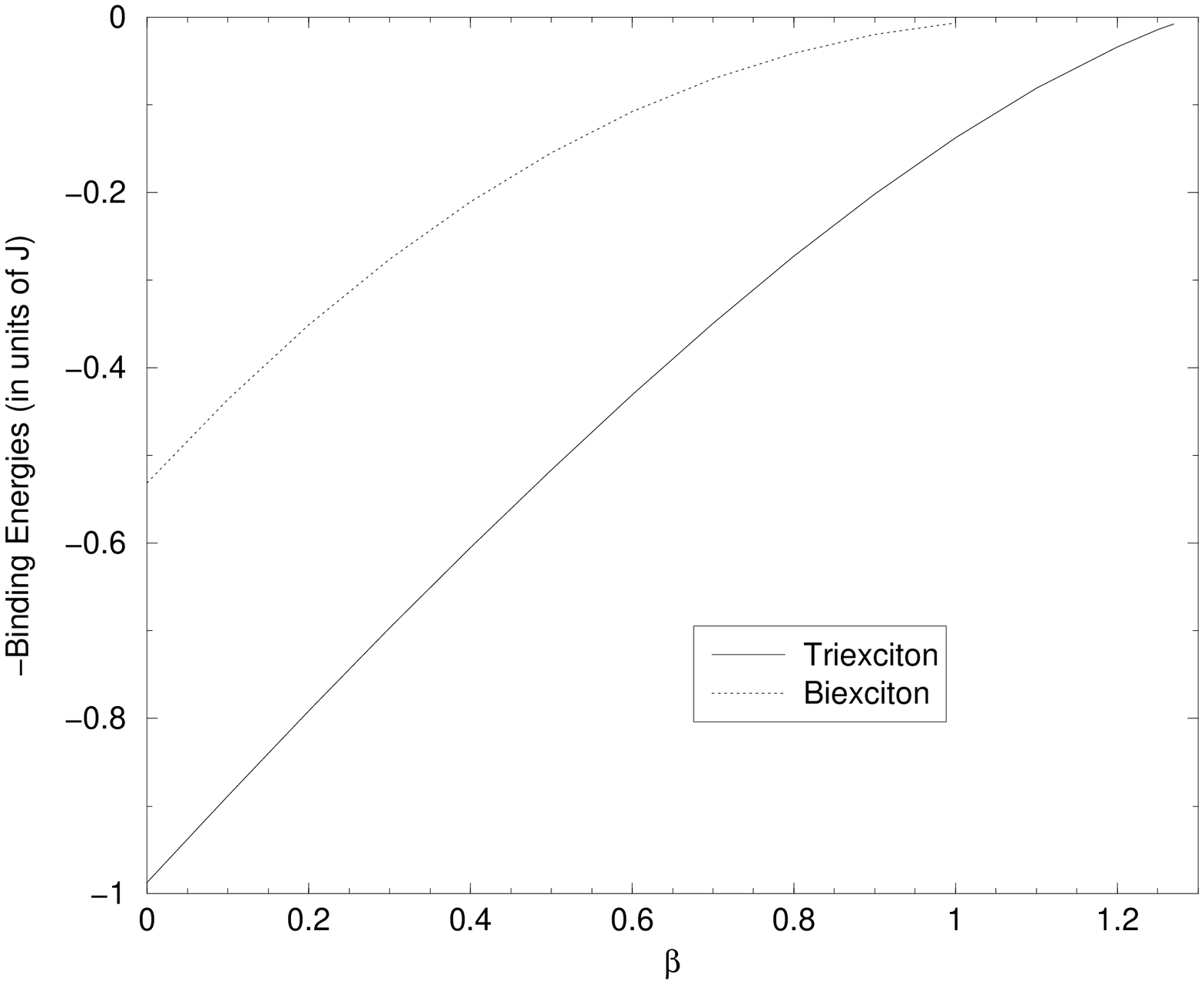,width=8cm,angle=-90}}
\caption{Negative of the binding energies for the bi- and tri-excitons for
$\Delta=3.33$ in function of $\beta=\frac{\alpha}{J}$ ($Q=0$)}
\label{bindingenergy}
\end{figure}

The binding energy of the biexciton, $E_{b}$, is then given by the following
equation
\begin{equation}
E_{b}(\alpha)=J(Q)[(e^{-\mu}+e^{\mu})-2]
\end{equation}The study of this expression shows that the biexciton
binding energy decreases when $\alpha$ increases (see figure \ref{bindingenergy}). This
behaviour is not surprising since the $\alpha$ term mixed together $R$
and $L$ particles; by doing so, configurations with two unlike
particles, $RL$ or $LR$, appear in the wave function which increase
its energy and then reduce the binding energy with respect to the case for $\alpha =0$. The
critical value 
$\alpha_{c}$ for which the binding energy becomes zero is given by the
solution of the equation $e^{\mu}+e^{-\mu}=2$,

\begin{equation}
\label{alphac}
\beta_{c}=\frac{\alpha_{c}}{J(Q)}=\frac{1}{2}\frac{(1-\lambda^{2})^{2}}{\lambda^{2}}
\end{equation}The critical value, $\alpha_{c}$, increases when $\lambda$ decreases, to
diverge for $\lambda=0$: for a very large binding energy, $\alpha$ has
no effect on the bound states anymore. Another interesting quantity
along this critical line, is the critical value $V_{c}$

\begin{equation}
\frac{1}{\lambda_{c}}=\frac{V_{c}}{J(Q)}=\frac{2}{\sqrt{4+2\beta}-\sqrt{2\beta}}
\end{equation}From this equation, we can see without any surprise,
than $V_{c}$ - the critical value above which one gets biexcitonic
states - continuously increases when $\beta$ increases. The attraction
between excitons must be stronger when the mixing between $R$ and $L$
bosons becomes more important.
 
We analyze now the wave function and the resulting transition
moment associated with the one exciton-biexciton transition at $Q=0$. We proceed
as in the previous case; we write
$A_{+}=a_{+}e^{-\frac{\delta_{+}}{2}}$,
$B_{+}=a_{+}e^{\frac{\delta_{+}}{2}}$ and
$A_{-}=a_{-}e^{-\frac{\delta_{-}}{2}}$,
$B_{-}=a_{-}e^{\frac{\delta_{-}}{2}}$ and get with equations
(\ref{PBC}) and (\ref{set2})

\begin{equation}
\label{wf2}
\varphi_{\sigma_{1}\sigma_{2}}={\cal N}_{\alpha}\left(\cosh( \mu [n-\frac{N}{2}])+(-1)^{\delta_{\sigma_{1},\sigma_{2}}}e^{(\mu+\ln\lambda)(N-3)}\cosh([\mu+2\ln\lambda][n-\frac{N}{2}])\right)
\end{equation}where again $1<n<N-2$. ${\cal N}_{\alpha}$ is the
normalization constant which could be easily calculated if
necessary. The first term of (\ref{wf2}) comes from the $\varphi_{+}$
component, the second term from the $\varphi_{-}$ component. For
$\alpha=0$, $\mu=-\ln (\lambda)$ and the previous result is
recovered. For $\alpha\neq 0$, the weight of the $\varphi_{-}$
component decreases smoothly in the wave function. Consequently,
the a priori unfavourable configurations $RL$ or $LR$ appear in
the biexciton wave function, on the contrary to the previous case
without the $\alpha$ interaction.

The transition moment of the one exciton-biexciton transition at $Q=0$
can be easily calculated with the expression of the wave function
(\ref{wf2}), but contains no new information compared to what was already
pointed out for $\alpha=0$. The results obtained in the previous subsection remain
valid in this case. The only difference comes from the fact that
the $\varphi_{-}$ component is non active in linear
absorption. Consequently, for the same binding energy, if we assume
$\mid \vec{m_{cp}} \mid= \mid \vec{m_{oc}} \mid$, the biexcitonic
peak would be a bit more intense for $J$ ($H$) aggregates or CT
solids than for CP. For instance, for $\alpha=0$, we get $\mid \vec{M_{cp}}
\mid^{2}=\frac{\mid \vec{M_{oc}} \mid^{2}}{2}$; for $\alpha\neq 0$,
the $\varphi_{+}$ becomes more important
than the $\varphi_{-}$ component in the wave function: the intensity
of the one exciton-biexciton transition then increases.

Now, let us take reasonable parameters
for Conjugated Polymers. Extracting from experimental data (we take
roughly the binding energy of the $1B_{u}^{+}$ exciton state observed in
Polydiacetylene compounds), we get $4J=0.5eV$ \cite{sylvie}. By
calculating the Coulomb
interaction between two HCB using the Mataga potential (for instance) \cite{revue} and
for a reasonable lattice constant ($a=1.5\AA$), we get $V=0.5eV.$ The
$\alpha$ term may be evaluated from the
energy difference between the $1B_{u}^{+}$ exciton and the
$nA_{g}^{-}$ "ionic" exciton. These states are represented here, in an
effective way, by $\mid +,0>$ and $\mid -,0>$ respectively (see
eq. \ref{1exciton}). This quantity could be estimated, for instance,
from exact resolutions of the PPP Hamiltonian for small clusters or, in
a better way, from electroabsorption experiments. There, the state $\mid
+,0>$ shows a red-shift with no change of the spectral line and a
quadratic dependence on the applied field. This quadratic Stark
effect of the exciton can be reasonably well reproduced with the
state $\mid -,0>$ only, neglecting, in particular, the
continuum effects \cite{sylvie}. Within this assumption, it is then very easy to estimate
$\alpha$. We take here a reasonable value, $\alpha=0.2eV$. With
this choice of
parameters, we get a binding energy of $0.12eV$ for the biexcitons
which is in accordance with experimental results
\cite{pa,tpa}. By the way, it is easy within our model to get such a value
for the binding energy for other values of $\alpha$ or/and $J$ by changing $V$,
which is more flexible.

\begin{figure}
\centerline{\psfig{figure=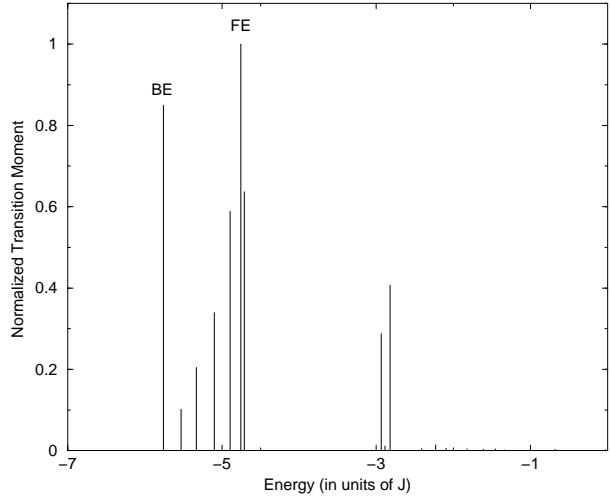,width=8cm,angle=-90}}
\caption{Normalized transition moment for the one exciton - 2 exciton
transitions for a finite cluster (N=10),
$\Delta=3.33$ and $\beta=0.2$. BE (Bound Excitons) is the biexcitonic
peak; FE (Free Excitons) is the more intense peak due to two free
excitons. At the thermodynamic limit, only FE will survive, slightly
shifted toward the low energies.}
\label{finitesize}
\end{figure}

In figure \ref{finitesize}, a calculation for a finite cluster of ten
sites is shown. The parameters are $\Delta=3.33$ and $\beta=0.2$ which
give a typical spectrum. BE is for the peak associated with the
BiExciton and, FE, for the peak associated with the more intense two Free Exciton state
(even if such concepts are not well defined for finite
clusters). The BE peak is a bit smaller than the FE one, which is in
accordance with the results of ref. \cite{mazumdar} from exact
calculations using the Extended Hubbard Hamiltonian, a short version
of the PPP Hamiltonian. At the thermodynamics limit,
only the FE peaks will survive, slightly shifted toward the low
energies. 

To conclude with the two particle case, we may summarize our main
results. For $\alpha=0$, biexcitons exist if $0<\lambda<1$; the
transition moment of the one exciton-biexciton transition shows inverse
behaviour compared to the ground-state-one exciton transition: it
decreases when the biexciton binding energy increases and, for
sufficiently large $N$, is independent of the system size. For $\alpha
\ne 0$, the conclusions aforementioned remain valid at the condition
that $\alpha<\alpha_{c}$ for $J$ and $V$ fixed. Above this critical
value, biexcitons fall in the continuum. Finite size cluster
calculations give qualitatively similar results as other calculations performed
with more complicated Hamiltonians.

\section{three (and more) exciton states}

For more than two particles, the question about the integrability of
the Hamiltonian (\ref{hamiltonian}) becomes important. To answer this
question, it is sufficient to analyze carefully the case with only two
particles \cite{sutherland,izyumov}. For that purpose, let us first
rewrite the Hamiltonian (\ref{hamiltonian}) in a more convenient way
by introducing two new operators

\begin{equation}
\begin{array}{c}
B^{\dagger}_{n}=\frac{1}{\sqrt{2}}(R^{\dagger}_{n}+L^{\dagger}_{n})\\
A^{\dagger}_{n}=\frac{1}{\sqrt{2}}(R^{\dagger}_{n}-L^{\dagger}_{n})
\end{array}
\end{equation}Then, the Hamiltonian becomes diagonal in $\alpha$
\begin{equation}
\label{hamiltonian2}
\begin{array}{c}
H=(\omega_{0}-\alpha)\sum_{n}B^{\dagger}_{n}B^{\quad}_{n}+(\omega_{0}+\alpha)\sum_{n}A^{\dagger}_{n}A^{\quad}_{n}
-J\sum_{n}(B^{\dagger}_{n}B^{\quad}_{n+1}+h.c.)\\-J\sum_{n}(A^{\dagger}_{n}A^{\quad}_{n+1}+h.c.)
-V\sum_{n}(A^{\dagger}_{n}B^{\quad}_{n}+B^{\dagger}_{n}A^{\quad}_{n})(A^{\dagger}_{n+2}B^{\quad}_{n+2}+B^{\dagger}_{n+2}A^{\quad}_{n+2})
\end{array}
\end{equation}To illustrate this new picture, note that the states with one
exciton (\ref{1exciton}) are the Bloch functions of one $B$ or one $A$
particle.

Let us now consider the case with two like particles ($BB$ or
$AA$) located in $n_{1}$ and $n_{2}$ ($n_{1}<n_{2}-1$). Because the
last term of (\ref{hamiltonian2}) exchanges the color ($B$ or $A$) of the
two particles, the corresponding amplitude in the wave function is
written as
\begin{equation}
\begin{array}{c}
\Psi(n_{1},n_{2})={\cal C}(B,k_{1};B,k_{2})e^{ik_{1}n_{1}+ik_{2}n_{2}}+{\cal C}(B,k_{2};B,k_{1})e^{ik_{2}n_{1}+ik_{1}n_{2}}+\\{\cal C}(A,k_{1}^{'};A,k_{2}^{'})e^{ik_{1}^{'}n_{1}+ik_{2}^{'}n_{2}}+{\cal C}(A,k_{2}^{'};A,k_{1}^{'})e^{ik_{2}^{'}n_{1}+ik_{1}^{'}n_{2}}
\end{array}
\end{equation}the constants ${\cal C}$ being complex numbers
determined in the previous section by solving the Schr\"odinger equation.

The system is invariant by translation, hence the total momentum is
conserved: $k_{1}+k_{2}=k_{1}^{'}+k_{2}^{'}=Q$. We then write $k_{1}=Q-q$,
$k_{2}=Q+q$ and $k_{1}^{'}=Q-k$, $k_{2}^{'}=Q+k$.

Now, let us do a scattering 'experiment'. Starting, for instance, with two $B$ particles, after the scattering
processes we get two $A$ particles. The collision being elastic, the
energy (\ref{energie}) must be conserved
\begin{equation}
-2\alpha-J(Q)(e^{q}+e^{-q})=2\alpha-J(Q)(e^{k}+e^{-k})
\end{equation}

Depending on the $\alpha$ value, we can make some conclusions.
\begin{itemize}

\item If $\alpha \ne 0$, $q \ne k$ (see eq. \ref{wf2} for
instance). Hence, there is a change in the momenta during the scattering
processes. In other words, diffractive processes are involved during
the scattering  which is sufficient to conclude that the model defined by (\ref{hamiltonian}) (or (\ref{hamiltonian2})) is non-integrable \cite{sutherland}.

\item If $\alpha=0$, $q=k$. There is no diffractive processes
anymore; then, as we look for the lowest bound-states, we will see
that the model (1) may be expressed, in that case, as an effective
spin-1/2 {\sl XXZ} Hamiltonian which is well known to be integrable
\cite{sutherland,izyumov,takahashi,gaudin}. Rigorously, to conclude
about the integrability of this model in the general case, the study
of the two particles $S$ matrix has to be done \cite{izyumov}. We
will not attempt such study here.
\end{itemize}

In the following two subsections, we will study both cases, starting
with the simplest one where $\alpha=0$.

\subsection{$\alpha=0$}

For $\alpha=0$, by interacting with each other, particles do not
exchange their color; the interaction is reduced to an exchange
of momenta alone. In such case, it is better to consider the
Hamiltonian (\ref{hamiltonian}) expressed in terms of $R$ and $L$
particles. 

Since $R$ and $L$ particles are HCB, an {\sl n}-electron configuration is
subdivided into {\sl n} separated regions. Each particle stays  
in its own area of the chain; within optical terminology, only reflections
are allowed. The Hamiltonian is studied independently for each
configuration of color which corresponds to a configuration of
pair-interactions, attractive or repulsive, depending on the nature of
the neighbouring particles. In the
general case, the problem is very complex. However, the interesting
case for us is given by the configurations where all the particles have
the same color. In this case, the model can be written in the form
of a {\sl XXZ} ferromagnetic Hamiltonian for spin-1/2 which has been solved
exactly by using the Bethe-Ansatz \cite{bethe,izyumov,takahashi,gaudin}. 

First, our goal is to show the equivalence of our model
(\ref{hamiltonian}) in the low energy sector and the {\sl XXZ}
model. For simplicity, we consider here particles without
extension ($r=0$). Of course, it is possible to work with our
defined $R$ and $L$ particles, but the comparison with the {\sl XXZ}
model would then claimed some additional notations unnecessary for our
purpose. By the way, in the second subsection of this section, we will
show more formally the equivalence between the two models working with
the extended particles.

Second, since we consider the simplest case with {\sl n}-particles like - for
instance, the states with {\sl n} {\sl R} particles - the Hamiltonian of the
system may be expressed in terms of spin-one-half
operators: $S^{z}_{n}=+\frac{1}{2}$ being for a site $n$ without any
particle and $S^{z}_{n}=-\frac{1}{2}$, for a site occupied by a $R$
particle (an exciton). The model becomes then equivalent to the
spin-1/2 $XXZ$ model with an 'attractive' anisotropic term \cite{frenkel}

\begin{equation}
\label{XXZ}
H=2\omega_{0}\sum_{n}S^{z}_{n}-2J\sum_{n}\frac{1}{2}(S^{+}_{n}S^{-}_{n+1}+S^{-}_{n}S^{+}_{n+1})-V\sum_{n}(S^{z}_{n}-\frac{1}{2})(S^{z}_{n+1}-\frac{1}{2})
\end{equation}The Ground state of (\ref{XXZ}) is the state
with only up-spins, $|GS>=|...\uparrow \uparrow \uparrow \uparrow
\uparrow ...>$. $S^{-}_{n}$ creates a down-spin (a $R$ particle or exciton)
localized at site $n$, $S^{-}_{n}|GS>=|...\uparrow \uparrow \uparrow
\downarrow_{n} \uparrow \uparrow \uparrow...>$. We introduce here, as
usual, the parameter of anisotropy $\Delta=\frac{V}{2J}$
\cite{izyumov,takahashi,gaudin}.

This model has been subject of intense studies since several decades,
and many results are known about it
\cite{izyumov,takahashi,gaudin}. Concerning the bound states, or more
precisely, the so-called "string-states", their properties depend on
the anisotropic parameter $\Delta$. Within the "string
hypothesis", the following results were
found\cite{izyumov,takahashi,gaudin,ovchinnikov}.
\begin{itemize}
\item If $\Delta>1$, the bound
states of $n$-spins, here $n$-excitons, exist without
any restriction on $n$
\cite{izyumov,takahashi,gaudin,ovchinnikov}. The energy of a
$n$-excitonic string is then given by the following formula first appeared
in \cite{ovchinnikov}
\begin{equation}
\label{strings}
E_{s}^{(n)}=n\omega_{0}-nV-J\sinh \gamma \frac{\cos Q-\cosh n \gamma}{\sinh n \gamma}
\end{equation}with $\cosh \gamma=\Delta$ and $Q$ being the
wave number of the n-string center of mass. For $n=2$, the energy
(\ref{energiebiexciton}) is of course recovered.

\item If $0<\Delta <1$, the $n$-excitonic strings may exist but with
a strong restriction on $n$
\begin{equation}
(n-1)\Theta< \pi
\end{equation}with $\cos \Theta = \Delta$,
$0<\Theta<\frac{\pi}{2}$. Moreover, these excitonic strings fall in
the continuum spectrum.

\end{itemize}

We have seen in the previous section, that biexcitons exist if
$\Delta>1$. In this case, from the above conclusions about the
{\sl XXZ}-model, we can say that $n$-excitonic strings will also exist and
that their binding energies will increase with $n$
(\ref{strings}); this result is in agreement with numerical studies on
small clusters \cite{string}. However, as we have shown in detail for the biexcitonic
case, it is not easy to observe such states
experimentally, since the particles are tightly bounded inside the complex, resulting in a lack of oscillator strength. For $0<\Delta<1$, excitonic strings may still  exist
but with an energy falling in the continuum. In this limit, the excitons
are less bounded and, therefore, it could be interesting to study the
signatures of these states under the influence of an electric field. We
leave this question to further work.

\subsection{$\alpha \ne 0$}

For $\alpha \ne 0$, the interactions between particles involve both
exchange of momenta and exchange of color. By a simple argument,
showing the appearance of diffractive terms in the two particles
scattering case, the non-integrability of the model
(\ref{hamiltonian2}) has been shown. Therefore, to study {\sl n}-excitonic
strings with $n>2$ one needs to do some approximations. In the
following, we propose a trial wave function for the three particle
case build from the Bethe Ansatz
solution of the {\sl XXZ}-Heisenberg model within the string hypothesis.

Let us consider the expression (\ref{hamiltonian2}) of our model
expressed in terms of $A$ and $B$ particles. For the three particle
case, eight configurations have to be distinguished depending on the
color of the particles and on their relative positions; we denote these
configurations as $\sigma_{1}\sigma_{2}\sigma_{3}$, where $\sigma_{i}$ is
for the color ($A$ or $B$). These configurations are separated into
two disconnected channels: on one hand, we have $AAA$, $BBA$, $BAB$
and $ABB$; on the other hand, $BBB$, $AAB$, $ABA$ and $BAA$. We
consider only the latter channel which is the lowest in energy.

The wave function for the
triexciton is expressed as a linear combination
\begin{equation}
\label{wf3}
\mid \psi_{T}>=a_{0}\mid \psi_{0}>+a_{l}\mid \psi_{l}>+a_{c}\mid \psi_{c}>+a_{r}\mid \psi_{r}>
\end{equation}where
$x=0,r,c,l$ stands for the $BBB$, $AAB$, $ABA$ and $BAA$ states respectively
($r,c,l$ for {\em right}, {\em center} and {\em left} refer to the
position of the $B$ particle in the three particles complex). $a_{x}$
are the coefficients to determine. Every component $\mid \Psi_{x}>$
is assumed to have the Bethe ansatz structure
\begin{equation}
\label{amplitude}
\mid \Psi_{x}>=\sum_{1\leq n_{1}<n_{2}<n_{3}\leq
N-1}\psi_{x}(n_{1},n_{2},n_{3})\mid n_{1},n_{2},n_{3}>_{x}
\end{equation}where, more precisely, the particles are again separated
by one bond, at least, and with
\begin{equation}
\label{BA}
\psi_{x}(n_{1},n_{2},n_{3})=\sum_{{\cal P}\in
S_{3}}e^{i\sum_{j=1,3}k^{x}_{{\cal
P}_{j}}n_{j}+\frac{i}{2}\sum_{l<j}\theta^{x}_{{\cal P}_{l}{\cal P}_{j}}}
\end{equation}The $k_{i}^{x}$ are the quasi-momenta and the
$\theta_{ij}^{x}$ the phase terms \cite{izyumov,takahashi,gaudin}. The summation runs over the $3!$
permutations of the indices. We apply periodic boundary conditions
to the wave functions,
$\psi_{x}(n_{2},n_{3},n_{1}+N)=\psi_{x}(n_{1},n_{2},n_{3})$, and we get the
well known equations \cite{bethe}

\begin{equation}
e^{ik^{x}_{{\cal P}_{j}}N+i\sum_{a}\theta^{x}_{{\cal P}_{a}{\cal
P}_{j}}}=1
\end{equation}

Since we are looking for excitonic complexes, we are interested in complex
solutions of these equations. Following Ovchinnikov \cite{ovchinnikov}, we assume that the only relevant
phase terms are $\theta_{12}^{x}$ and $\theta_{23}^{x}$, then
\begin{equation}
\left \{ \begin{array}{l}
Nk_{1}^{x}=2\pi \lambda_{1}^{x}+\theta_{12}^{x}\\
Nk_{2}^{x}=2\pi \lambda_{2}^{x}-\theta_{12}^{x}+\theta_{23}^{x}\\
Nk_{3}^{x}=2\pi \lambda_{3}^{x}-\theta_{23}^{x}
\end{array} \right.
\end{equation}where $\lambda_{i}^{x}$ are integers between $0$ and
$N-1$. It follows
\begin{equation}
\left \{ \begin{array}{l}
N \mbox{Im}k_{1}^{x}=\mbox{Im}\theta_{12}^{x}\\
N\mbox{Im}k_{2}^{x}=\mbox{Im}(\theta_{23}^{x}-\theta_{12}^{x})\\
N\mbox{Im}k_{3}^{x}=-\mbox{Im}\theta_{23}^{x}
\end{array} \right. => \left \{ \begin{array}{l}
2\mbox{Im}\theta_{12}^{x}=N(\mbox{Im}k_{1}^{x}-\mbox{Im}k_{2}^{x})\\
2\mbox{Im}\theta_{23}^{x}=N(\mbox{Im}k_{2}^{x}-\mbox{Im}k_{3}^{x})
\end{array} \right.
\end{equation}

Next, we assume $\theta_{12}^{x}=\mbox{Im}\theta_{12}^{x}<0$ and
$\theta_{23}^{x}=\mbox{Im}\theta_{23}^{x}<0$ without loss of generality; then, at the thermodynamic limit, among
the summation (\ref{BA}) only one term remains with an
exponential accuracy in $N$

\begin{equation}
\label{wfstring}
\psi_{x}(n_{1},n_{2},n_{3})=e^{ik^{x}_{1}n_{1}+ik^{x}_{2}n_{2}+ik_{3}^{x}n_{3}+\theta^{x}_{12}+\theta^{x}_{23}}
\end{equation}We may notice, that the used assumption about the phase
terms $\theta_{ij}^{x}$, is rigorously correct in the case without
$\alpha$ \cite{ovchinnikov}.

The wave function (\ref{wf3}) with the simplification
(\ref{wfstring}), must satisfy the following constraints due to the $V$
term, in order to be an eigenfunction of (\ref{hamiltonian2}):
\begin{itemize}
\item for $n_{2}=n_{1}+2$, ($\forall n_{3}$)
\begin{equation}
\label{constraints1}
\begin{array}{l}
a_{0}J(\Psi_{0}(n_{1}+1,n_{1}+2,n_{3})+\Psi_{0}(n_{1},n_{1}+1,n_{3}))-a_{l}V\Psi_{l}(n_{1},n_{1}+2,n_{3})=0
\\
a_{l}J(\Psi_{l}(n_{1}+1,n_{1}+2,n_{3})+\Psi_{l}(n_{1},n_{1}+1,n_{3}))-a_{0}V\Psi_{0}(n_{1},n_{1}+2,n_{3})=0
\\
a_{c}J(\Psi_{c}(n_{1}+1,n_{1}+2,n_{3})+\Psi_{c}(n_{1},n_{1}+1,n_{3}))-a_{r}V\Psi_{r}(n_{1},n_{1}+2,n_{3})=0
\\
a_{r}J(\Psi_{r}(n_{1}+1,n_{1}+2,n_{3})+\Psi_{r}(n_{1},n_{1}+1,n_{3}))-a_{c}V\Psi_{c}(n_{1},n_{1}+2,n_{3})=0
\end{array}
\end{equation}
\item for $n_{3}=n_{2}+2$, ($\forall n_{1}$)
\begin{equation}
\label{constraints2}
\begin{array}{l}
a_{0}J(\Psi_{0}(n_{1},n_{2}+1,n_{2}+2)+\Psi_{0}(n_{1},n_{2},n_{2}+1))-a_{r}V\Psi_{r}(n_{1},n_{2},n_{2}+2)=0
\\
a_{r}J(\Psi_{r}(n_{1},n_{2}+1,n_{2}+2)+\Psi_{r}(n_{1},n_{2},n_{2}+1))-a_{0}V\Psi_{0}(n_{1},n_{2},n_{2}+2)=0
\\
a_{c}J(\Psi_{c}(n_{1},n_{2}+1,n_{2}+2)+\Psi_{c}(n_{1},n_{2},n_{2}+1))-a_{l}V\Psi_{l}(n_{1},n_{2},n_{2}+2)=0
\\
a_{l}J(\Psi_{l}(n_{1},n_{2}+1,n_{2}+2)+\Psi_{l}(n_{1},n_{2},n_{2}+1))-a_{c}V\Psi_{c}(n_{1},n_{2},n_{2}+2)=0
\end{array}
\end{equation}
\end{itemize}

Moreover, by symmetry, we have $k^{r}_{i}=k^{l}_{i}$,
$\theta^{r}_{i,i+1}=\theta^{l}_{i,i+1}, \forall i$ and
$a_{r}=a_{l}$. Then, the previous set of equations together with the
expression (\ref{wfstring}) gives 
\begin{equation}
\label{system}
\left \{ \begin{array}{l}
J[e^{i(k_{1}^{0}+k_{2}^{0})}+1]-Ve^{ik_{2}^{0}}\frac{V}{J}\frac{e^{ik_{2}^{l}}}{e^{i(k_{1}^{l}+k_{2}^{l})}+1}=0\\
J[e^{i(k_{2}^{0}+k_{3}^{0})}+1]-Ve^{ik_{3}^{0}}\frac{V}{J}\frac{e^{ik_{3}^{l}}}{e^{i(k_{2}^{l}+k_{3}^{l})}+1}=0\\
J[e^{i(k_{1}^{l}+k_{2}^{l})}+1]-Ve^{ik_{2}^{l}}\frac{V}{J}\frac{e^{ik_{2}^{c}}}{e^{i(k_{1}^{c}+k_{2}^{c})}+1}=0\\
J[e^{i(k_{2}^{l}+k_{3}^{l})}+1]-Ve^{ik_{3}^{l}}\frac{V}{J}\frac{e^{ik_{3}^{c}}}{e^{i(k_{2}^{c}+k_{3}^{c})}+1}=0
\end{array} \right.
\end{equation}to which we add the momentum conservation law
\begin{equation}
\label{Q}
k_{1}^{x}+k_{2}^{x}+k_{3}^{x}=Q
\end{equation}$Q$ being the momentum associated with the motion of the
center of mass of the
excitonic complex. This system of equations is obviously
non-soluble in the general case. Hence, we add an additional simplification to the trial
wave function.

If $\alpha=0$, $k_{i}^{x}=k_{i}^{0}$, and therefore
$a_{x}=\frac{1}{2}, \forall x$; it is easy to verify
that the system (\ref{system}) becomes equivalent, in that case, to the
corresponding equations for the {\sl XXZ} Heisenberg model \cite{ovchinnikov} as we already
pointed out in the previous subsection. Then the system (\ref{system})
leads to the energy (\ref{strings}), for $\Delta>1$. Within the
string hypothesis, $k_{1}^{0}$ and $k_{3}^{0}$ are complex conjugate,
$k_{2}^{0}$, a purely real quantity.

Enlightened by the exact results for $\alpha=0$ \cite{ovchinnikov}, we assume for
$\alpha \ne 0$
\begin{equation}
\label{ansatz}
\left \{ \begin{array}{l}
k_{1}^{l}=k_{1}^{0}+\epsilon \\
k_{2}^{l}=k_{2}^{0}\\
k_{3}^{l}=k_{3}^{0}-\epsilon
\end{array} \right.
\end{equation}where $\epsilon$, a real number, is given by the following equation
obtained from the energy conservation law
\begin{equation}
\label{epsilon}
-3
\alpha-\sum_{i=1}^{3}(e^{ik_{i}^{0}}+e^{-ik_{i}^{0}})=\alpha-\sum_{i=1}^{3}(e^{ik_{i}^{l}}+e^{-ik_{l}^{l}})
\end{equation}The left-hand-side is the energy of the $BBB$
component, the right-hand-side, the energy of the three other components.

With this assumption on the momenta $k_{i}^{x}$, the solutions stay very close to
the Bethe Ansatz form; the binding
between the two bordered particles of the complex (with momenta $k_{1}^{x}$ and $k_{3}^{x}$),
ensures the cohesion of the three particle bound states. The set of equations composed
by the two first equations of (\ref{system}) and equations (\ref{Q}),
(\ref{ansatz}) and (\ref{epsilon}) are then solved numerically.

It is important to realize that our approximation is
variational. This important statement could be checked in the following way. One
starts with the trial wave function given by (\ref{wf3}) where the
components $\mid \Psi_{x}>$ are assumed to be of the form defined by the
equations (\ref{amplitude}), (\ref{wfstring}) and
(\ref{ansatz}) together with the restrictions (\ref{Q}) and
(\ref{epsilon}). The coefficients
$a_{x}$ of (\ref{wf3}), and their complex conjugates, are determined variationally by minimizing the functional
\begin{equation}
{\cal F}(\{ a_{x},a_{x}^{*}\})=< \Psi_{T}\mid H \mid
\Psi_{T}>-E<\Psi_{T}\mid \Psi_{T}>
\end{equation}where $-E$ is the Lagrange parameter for the
normalization constraint of the wave function. We get
\begin{equation}
\frac{\partial {\cal F}}{\partial a_{0}^{*}}=0 \Rightarrow
a_{0}<\Psi_{0}\mid H \mid \Psi_{0}>+a_{r}<\Psi_{0}\mid H \mid
\Psi_{r}>+a_{l}<\Psi_{0}\mid H \mid \Psi_{l}>=a_{0}E<\Psi_{0}\mid
\Psi_{0}>
\end{equation}and similar equations for all the other minimizations. At
the end of the day, we obtain nothing else than the usual secular
equations and one way to solve them, following A. Bethe \cite{bethe},
is to impose the constraints given by the sets of equations
(\ref{constraints1}) and (\ref{constraints2}). Our procedure is then
variational and the results found are an upper bound for the problem.

Typical results are shown in figure (\ref{bindingenergy}). The binding
energy of
the biexciton got in the previous section and the one of the
triexciton, obtained with our variational calculation, are represented
for $Q=0$ and $\Delta=3.33$ as a function of $\beta=\frac{\alpha}{J}$. The
triexciton binding energy is defined with respect to the continuum of
one biexciton plus one free exciton which is the lowest in energy. The
following remarks can be made. (i) The triexciton binding energy for
$\alpha=0$ is the exact one. (ii) The binding energy of the
triexciton is larger than the one of the biexciton. (iii) A critical
value of $\alpha$ exists also for the triexciton; it is
larger than $\alpha_{c}$ (\ref{alphac}) for the biexciton. It follows
that triexcitons may exist without biexcitons (in the sense that
$E_{b}>0$). We may suspect the same behaviour for the binding energy
of n-string for any $n$: the binding energy and the critical $\alpha$
may be larger for a n-string than for a (n-1)-string in accordance
with the results for the {\sl XXZ} model. (iv) Since our result is
variational, we get an upper bound for the triexciton binding energy
so that our conclusions are qualitatively correct. 

To illustrate the behaviour of the binding energy of a {\sl n}-string at
strong $\alpha$, we study this limit perturbatively. For $\alpha>> V$, the Hilbert space of (\ref{hamiltonian2}) is
naturally separated in two subspaces: the first, lower in energy, with
configurations without $A$ particles and, the second, higher in energy,
with configurations with at least one $A$ particle. The first one is
associated with the projector $P_{0}$, the second one with the
projector $P_{\eta}$ ($P_{0}+P_{\eta}=\hat 1$). The Hamiltonian may be
written as
\begin{equation}
H=P_{0}HP_{0}+P_{\eta}HP_{\eta}+P_{0}HP_{\eta}+P_{\eta}HP_{0}
\end{equation}

By a Schrieffer-Wolf type of canonical transformation
\cite{schrieffer,oles}, we want to derive an effective Hamiltonian $\tilde
H=P_{0}HP_{0}+P_{\eta}HP_{\eta}+\hat W$, with $P_{0}\tilde H
P_{\eta}=0$, $\hat W$ being the effective term to be determined. The study of the n-excitonic strings will then be
performed by diagonalizing the projection of this model into the space
without $A$ particle, $H_{eff}=P_{0}\tilde H P_{0}$. Introducing the
unitary transformation $U=e^{S}$ (with $S^{\dagger}=-S$), and
following the method described in detail in \cite{oles}, we found the
simple effective Hamiltonian valid in the strong coupling limit where
$\alpha>> V$, at second order in $\frac{V}{\alpha}$

\begin{equation}
H_{eff}=(\omega_{0}-\alpha)\sum_{n}B^{\dagger}_{n}B^{\quad}_{n}-2J\sum_{n}\frac{1}{2}(B^{\dagger}_{n}B^{
}_{n+1}+h.c.)-\frac{V^{2}}{4\alpha}\sum_{n}B^{\dagger}_{n}B^{
}_{n}B^{\dagger}_{n+2}B^{ }_{n+2}
\end{equation}We have recovered, once again, the spin-1/2 {\sl XXZ}
Hamiltonian (\ref{XXZ}). However, the anisotropic parameter is now given
by $\Delta=\frac{V^{2}}{8\alpha J}$, instead of $\frac{V}{2J}$, and the
interpretation of the spin operators is different:
$S^{z}_{n}=+\frac{1}{2}$ is still for a site without any particle but,
$S^{z}_{n}=-\frac{1}{2}$, is now for a site occupied by a $B$ particle
instead of a $R$ particle.

In the effective model, the attractive interaction is reduced by the $\alpha$ term and, for
the strong limit ($\alpha>>V$) where this derivation is valid, one
reaches a {\sl XX}-Heisenberg Hamiltonian which is well known to be equivalent
to a spinless free fermion system \cite{lieb}. Hence, in this limit,
the excitonic
strings do not exist which is consistent with the results of our
variational calculation for $\alpha>\alpha_{c}$. The states with free $B$ particles are then preferred.

\section{conclusion}
In this paper, we propose a simplified model to describe excitonic strings
in quasi-one dimensional organic compounds including organic Charge
Transfer solids, {\sl J}- and {\sl H}-aggregates and Conjugated Polymers. For all
these compounds, the excitonic states are characterized by small
radius. Hence, our model starts with the definition of Hard Core Boson
particles extended over one bond which describe, in an effective way,
the excitons in one dimensional organic compounds.

There are two possibilities in one dimension to place the electron and the hole, the two
elementary constituents of an exciton. Either the electron is on the
right or on the left of the hole; we introduce two kinds of HCB
which illustrate these two situations: the {\sl Right}-bosons and the
{\sl Left}-bosons. 

The proposed model in terms of these effective
particles contains four terms: the excitonic energy ($\omega_{0}$), which
separate in energy the space with $n$ particles from the space with
$n-1$ particles; the kinetic energy of the HCB ($J$); the local interaction
between the two species of bosons ($\alpha$), which illustrates the fact that
there exist effective interactions which can exchange the relative
position of the hole and the electron of an exciton; and last, the interaction between two excitons which are attractive between
particles of the same species and repulsive between particles of opposite
species (interaction of intensity $V$). The kinetic energy and the
$\alpha$-interaction combined with the repulsive two body
interaction act against the attractive two body interaction - they tend to delocalize the excitons on the
contrary to the two particle attraction which tends to create bound
states. The cases with two and three particles are mainly studied in this
work. We summarize briefly our results considering the case where $Q$,
the momentum of the center of mass of the excitonic complexes, is zero.

The case with two particles is solved exactly. For $\alpha=0$,
biexcitonic states exist if $\frac{V}{2J}>1$. Then, the intensity of
the transition from the one exciton state decreases dramatically with
the increase of the binding energy, to saturate and become independent of
the system size. This behaviour makes the biexcitonic state
difficult to observe at the thermodynamic limit. For $\alpha \ne 0$,
a critical value $\alpha_{c}$ is found, above which biexcitonic states
do not exist. Otherwise, the behaviour of the oscillator strength
remains qualitatively unchanged.

For $\alpha=0$, the $n$ particle case can be solved in the low energy sector. Then, the model is
equivalent to the spin one half {\sl XXZ} Heisenberg model which is exactly
solvable in one dimension. For $\frac{V}{2J}>1$, it is well known that
$n$-excitonic strings exist with a binding energy which increases with
$n$ \cite{izyumov,takahashi,gaudin,ovchinnikov}. For $\alpha \ne
0$, the model is non-integrable. The three particle case is then
studied using a variational ansatz, which may be extended to a more
general case with $n$ particles. Again, a 3-string is found with a
binding energy larger than the one for the biexciton. We also found a
critical value for $\alpha$, larger than the $\alpha_{c}$ valid for
the 2-strings.

From numerical studies on small clusters
\cite{string}, it has been argued that {\sl n}-strings may exist for any value of {\sl n}
in organic compounds. Our work gives some confidence to this
statement. Indeed, with our HCB Hamiltonian, if biexcitons exist,
triexcitons will exist with a larger binding energy; moreover, if
$\alpha=0$, {\sl n}-strings will also exist with binding energies
which increase with {\sl n} and, we believe, from our present results,
that it will also be the case for $\alpha \ne 0$.

The next important question concerning the {\sl n}-string peaks in PA
and TPA experiments, is fully clarified for the biexciton (the most
important case) and confirms some conclusions of previous works
\cite{frenkel,jh,mazumdar,vektaris,spano1,spano2}. The intensity of the transition from the one
exciton to the biexciton  decreases with the binding energy and, for
sufficient binding energy, becomes independent of the system
size. Hence, the PA intensity to the biexciton is proportional to
$\sqrt{N}$ instead of $N$ for the two free exciton case. This property comes from the fact that the two excitons are tightly
bounded within the complex, rendering the 2-string states difficult
to observe for infinite systems (clean samples are needed) and more
accessible experimentally for small oligomers \cite{frenkel,klimov}.

In the case of Conjugated Polymers, we think some additional studies
are needed going in two different directions. First, in the context of
PA experiments, excitations extended over two molecules have been
invoked \cite{excitations,hsu}; extension of our work to two dimensions
or, at least, for two coupled polymers would be suitable. Second, Conjugated
Polymers are disordered systems with several possible sources of disorder which
result, in practice, in the vague definition of the so-called
conjugation length \cite{excitations}. As we have already mentioned,
since the transitions to biexcitons are much less intense than the ones
involving two free excitons, we may expect at first analyse - and it
was implicitly our point of view all along this manuscript - that a
sufficient disorder will render the biexcitons not observable. However, some relatively recent results about the two
interacting particle problem in a disordered medium \cite{tip} show very
unexpected behaviours which may have dramatic consequences for our
problem, and possibly in a way reverse to our intuition. In the light
of the intriguing results of \cite{tip}, the
effects of disorder on transition moments should be clarified for
a correct interpretation of PA and TPA results. We leave these
considerations to further works.

\begin{acknowledgements}
I would like to thank Prof. A.A. Ovchinnikov for his support and illuminating
discussions. This work would certainly not be developed in the same way
without his help. This work has been supported by the European
Commission through the TMR network contract ERBFNRX-CT96-0079 (QUCEX).
\end{acknowledgements}

\newpage

\begin{references}
\bibitem{sc}
M. Ueta, H. Kanzaki, H. Kobayashi, Y. Toyozawa and E. Hanamura,
Excitonic Processes in Solids (Springer, Berlin, 1986).
\bibitem{hanamura}
E. Hanamura, Solid State Comm. {\bf 12}, 951 (1973).
\bibitem{gale}
G.M. Gale and A. Mysyrowicz, Phys. Rev. Lett. {\bf 32}, 727 (1974).
\bibitem{string}
S. Mazumdar, F. Guo, K. Meissner, B. Fluegel, N. Peyghambarian,
M. Kuwata-Gonokami, Y. Sato, K. Ema, R. Shimano, T. Tokihiro, H. Ezaki
and E. Hanamura, J. Chem. Phys {\bf 104}, 9283 (1996).
\bibitem{frenkel}
H. Ezaki, T. Tokihiro and E. Hanamura, Phys. Rev. {\bf B 50}, 10 506
(1994).
\bibitem{jh}
A. Chakrabarti, A. Schmidt, V. Valencia, B. Fluegel, S. Mazumdar,
N. Armstrong and N. Peyghambarian, Phys. Rev. {\bf B 57}, R4206,
(1998).
\bibitem{revue}
D. Baeriswyl, D.K. Campbell and S. Mazumdar, in {\sl Conjugated
Conducting Polymers}, edited by H. Kiess (Springer-Verlag, Heidelberg, 1992), pp7-133.
\bibitem{excitations}
{\sl Primary photoexcitations in Conjugated Polymers: Molecular Exciton versus Semiconductor Band Model}, edited by N.S. Sariciftci
(World Scientific Publishing, Singapore, 1997).
\bibitem{pa}
M.B. Sinclair, D. McBranch, T.W. Hagler and A.J. Heeger,
Synth. Met. {\bf 49-50}, 593 (1992).
\bibitem{klimov}
V.I. Klimov, D.W. McBranch, N. Barashkov and J. Ferraris,
Phys. Rev. {\bf 58}, 7654 (1998).
\bibitem{tpa}
P.D. Townsend, W.S. Fann, S. Etemad, G.L. Baker, Z.G. Soos and
P.C.M. McWilliams, Chem. Phys. Lett. {\bf 180}, 485 (1991).  
\bibitem{abe}
V.A. Shakin and S. Abe, Phys. Rev. {\bf B 50}, 4306 (1994).
\bibitem{mazumdar}
F. Guo, M. Chandross and S. Mazumdar, Phys. Rev. Lett. {\bf 74}, 2086
(1995).
\bibitem{vektaris}
G. Vektaris, J. Chem. Phys. {\bf 101}, 3031 (1994).
\bibitem{spano1}
F.C. Spano, Chem. Phys. Lett. {\bf 234}, 29 (1995).
\bibitem{spano2}
F.C. Spano and E. Manas, J. Chem. Phys. {\bf 103}, 5939 (1995).
\bibitem{yu}
Z.G. Yu, R.T. Fu, C.Q. Wu, X. Sun and K. Nasu, Phys. Rev. {\bf B52},
4849 (1995).
\bibitem{gallagher}
F.B. Gallagher and F.C. Spano, Phys. Rev. {\bf B 53}, 3790 (1996).
\bibitem{davydov}
A.S. Davydov, Theory of Molecular Excitons, New York, 1971.
\bibitem{chandross}
M. Chandross, Y. Shimoi and S. Mazumdar, Phys. Rev. {\bf B 59}, 4822 (1999).
\bibitem{pleutin1}
S. Pleutin and J.L. Fave, J. Phys. Cond. Matt. {\bf 10}, 3941 (1998).
\bibitem{pleutin2}
S. Pleutin, in preparation.
\bibitem{simpson}
W.T. Simpson, J. Am. Chem. Soc. {\bf 77}, 6164 (1955).
\bibitem{rice}
M.J. Rice and Y.N. Gartstein, Phys. Rev. Lett. {\bf 73}, 2504 (1994).
\bibitem{sylvie}
A. Horvath, G. Weiser, C. Lapersonne-Meyer, M. Schott and S. Spagnoli,
Phys. Rev. {\bf B 53}, 13507 (1996).
\bibitem{bethe}
H. Bethe, Z. Physik {\bf 71}, 205 (1931).
\bibitem{sutherland}
B. Sutherland, Synth. Met. {\bf 84}, 11 (1997).
\bibitem{izyumov}
Y.A Izyumov and Y.N. Skryabin, "Statistical Mechanics of Magnetically
Ordered Systems", Consultants Bureau, New-York and London (1988).
\bibitem{takahashi}
M. Takahashi, in "Conformal Field Theories and Integrable Models",
Springer-Verlag, 1997 (Berlin).
\bibitem{gaudin}
M. Gaudin, "La fonction d'onde de Bethe", Masson (1983).
\bibitem{ovchinnikov}
A.A. Ovchinnikov, JETP Lett. {\bf 5}, 38 (1967).
\bibitem{schrieffer}
J.R. Schrieffer and P.A. Wolf, Phys. Rev. {\bf 149}, 491 (1966)
\bibitem{oles}
K.A. Chao, J. Spalek and A.M. Oles, J. Phys C: Solid State Phys {\bf
10}, L271 (1977).
\bibitem{lieb}
E. Lieb, T. Schultz and D. Mattis, Annals of Phys. {\bf 16}, 407
(1961).
\bibitem{hsu}
J.W.P. Hsu, M. Yan, T.M. Jedju, L.J. Rothberg and B.R. Hsieh,
Phys. Rev. {\bf B 49}, 712 (1994).
\bibitem{tip}
O.N. Dorokhov, JETP {\bf 71}, 360 (1990);
D.L. Shepelyansky, Phys. Rev. Lett. {\bf 73}, 2607 (1994).
\end{references}
\end{document}